\documentclass[prd,reprint,nofootinbib,superscriptaddress]{revtex4-2}
\usepackage{bm}
\usepackage{amsmath}
\usepackage{ulem}
\usepackage{changes}
\usepackage{latexsym}
\usepackage{amsfonts}
\usepackage{graphicx}
\usepackage{mathrsfs}
\usepackage{CJKutf8}
\usepackage{subfigure}
\usepackage{float}
\usepackage{orcidlink}
\usepackage{color}
\usepackage{hyperref}
\hypersetup{
	colorlinks=true,
	linkcolor=red,
	citecolor=blue,
}

\def\pa{\partial}
\def\n{\nonumber}
\def\l{\left}
\def\r{\right}

\begin{document}

\begin{CJK*}{UTF8}{gbsn}


\title{Extreme mass ratio inspirals in galaxies with dark matter halos}
\author{Ning Dai (戴宁) \orcidlink{0000-0002-0867-6764}}
\email{daining@hust.edu.cn}
\affiliation{School of Physics, Huazhong University of Science and Technology, Wuhan, Hubei 430074, China}
\author{Yungui Gong (龚云贵) \orcidlink{0000-0001-5065-2259}}
\email{Corresponding author. yggong@hust.edu.cn}
\affiliation{Institute of Fundamental Physics and Quantum Technology, Department of Physics, School of Physical Science and Technology, Ningbo University, Ningbo, Zhejiang 315211, China}
\affiliation{School of Physics, Huazhong University of Science and Technology, Wuhan, Hubei 430074, China}
\author{Yang Zhao (赵阳)
\orcidlink{0009-0003-7436-8668}}
\email{zhaoyangedu@hust.edu.cn}
\affiliation{School of Physics, Huazhong University of Science and Technology, Wuhan, Hubei 430074, China}
\author{Tong Jiang (江通)
\orcidlink{0000-0003-3986-1433}}
\email{jiangtong@hust.edu.cn}
\affiliation{College of Physics and Technology, Kunming University, Kunming 650214, China}
%


\begin{abstract}
Using an analytic, static, and spherically symmetric metric for a Schwarzschild black hole immersed in a dark matter (DM) halo with the Hernquist-type profile,
we derive analytic expressions for the orbital period and precession of eccentric extreme mass ratio inspirals (EMRIs) surrounded by DM halos,
and we show how the precession rates decrease and even undergo a prograde-to-retrograde precession transition if the density of DM halo is large enough.
The presence of local DM halos also retards the decrease of the semi-latus rectum and the eccentricity.
The orbital evolution of EMRIs immersed in DM halos is then calculated numerically by considering the combined effects of gravitational radiation reaction, dynamical friction, and accretion.
Comparing the number of orbital cycles accumulated over a one-year evolution for EMRIs with and without DM halos, 
we find that DM halos with compactness as small as $10^{-5}$ can be detected.
From the mismatch between gravitational waveforms of EMRIs with and without DM halos, 
we show that EMRIs in galaxies can be used to probe the existence of DM halos and detect the compactness of DM halos as small as $10^{-5}$.
Employing the Fisher information matrix method, 
we find that larger compactness and density values of DM halos help to reduce the estimation error of parameters and further break the degeneracy between the parameters.

\end{abstract}

\maketitle

\section{Introduction}

The first detection of gravitational wave (GW) from the merger of a black hole (BH) binary by the LIGO Scientific Collaboration and the Virgo Collaboration in 2015 \cite{LIGOScientific:2016aoc,LIGOScientific:2016emj} opened a new window for probing gravitational physics and fundamental physics.
Since then, tens of confirmed GW events have been detected by the ground-based GW observatories \cite{LIGOScientific:2018mvr,LIGOScientific:2020ibl,LIGOScientific:2021usb,KAGRA:2021vkt}.
The ground-based GW observatories are only sensitive to GWs in the frequency range of $10-10^3$ Hz.
The space-based GW observatories such as LISA \cite{Danzmann:1997hm,LISA:2017pwj,Colpi:2024xhw}, TianQin \cite{TianQin:2015yph}, and Taiji \cite{Hu:2017mde, Gong:2021gvw} will usher a new era in GW astronomy due to their unprecedented accuracy and their sensitive band of mHz \cite{Baibhav:2019rsa,Amaro-Seoane:2022rxf,LISA:2022kgy,Karnesis:2022vdp}.
One particularly interesting target of space-based GW detectors is the case of a stellar-mass compact object (SCO) inspiraling into
 a massive black hole (MBH)-the extreme mass ratio inspirals (EMRIs) \cite{Babak:2017tow}.
There are $10^5-10^6$ GW cycles in the detector band when the SCO inspirals deep inside the strong field region of the MBH,
and rich information about the spacetime geometry around the MBH is encoded in the GW waveform. 
Therefore, the observations of GWs emitted from EMRIs present us a pristine opportunity for the study of astrophysics, gravity in the strong and nonlinear regions, and the nature of BHs  \cite{Amaro-Seoane:2007osp,Berry:2019wgg,Seoane:2021kkk,Laghi:2021pqk,McGee:2018qwb}.

Although the property of 
dark matter (DM) is still a mystery in particle physics, there is a plethora of indirect evidence for the existence of 
DM in the Universe \cite{vandenBergh:1999sa,Rubin:1970zza,Rubin:1980zd,Begeman:1991iy,Persic:1995ru,Corbelli:1999af,Moustakas:2009na,Massey:2010hh,Ellis:2010kf,Challinor:2012ws}.
DM may cluster at the centers of galaxies and around BHs \cite{Sadeghian:2013laa,Navarro:1996gj,Gondolo:1999ef,Hernquist:1990be},
and affect the dynamics of binaries, and hence the GWs emitted from them.
Since EMRIs are believed to reside in stellar clusters and the centers of galaxies, DM may affect the dynamics of EMRIs and be detectable in
the observations of GWs from EMRIs.
In particular, those EMRIs
immersed in DM environment may be used to understand the astrophysical environment surrounding them,  
probably confirm the existence of DM and uncover the nature of DM \cite{Yunes:2011ws,Kocsis:2011dr,Eda:2013gg,Macedo:2013qea,Eda:2014kra,Barausse:2014tra,Barack:2018yly,Hannuksela:2018izj,Cardoso:2019rou,Yue:2019ozq,Annulli:2020ilw,Derdzinski:2020wlw,Zwick:2022dih,Dai:2021olt,Coogan:2021uqv,Kim:2022mdj}.

In the studies of DM effects, Newtonian approaches were usually applied
and the gravitational effects of DM on the dynamical evolution of EMRIs were modeled at the Newtonian level.
In Ref. \cite{Cardoso:2021wlq},
the authors generalized Einstein clusters \cite{Einstein:1939ms,Geralico:2012jt} to include horizons,
solved Einstein equations sourced by DM halo of Hernquist type \cite{Hernquist:1990be} with a MBH at its center
and obtained an analytical formula for the metric of galaxies harboring MBHs.
Exact solutions for the geometry of a MBH immersed in DM halos with different density distributions were then derived \cite{Konoplya:2022hbl,Jusufi:2022jxu,Shen:2023erj}.
With the fully relativistic formalism, it was found that the leading-order correction to the ringdown stage induced by the external matter and fluxes by orbiting particles is a gravitational redshift,
and the difference between the numbers of GW cycles accumulated by EMRIs with and without DM halos over one year before the innermost stable circular orbit (ISCO) can reach about $500$ \cite{Cardoso:2021wlq}.
In galaxies harboring MBHs, tidal forces, and geodesic deviation depend on the mass of the DM halo and the typical length scale of the galaxies \cite{Liu:2022lrg}.
Due to the gravitational pull of DM halos, the apsidal precession of the geodesic orbits for EMRIs is strongly affected and even prograde-to-retrograde drift can occur \cite{Destounis:2022obl}.
In prograde-to-retrograde orbital alterations, GWs show transient frequency phenomena around a critical non-precessing turning point.  
A fully relativistic formalism to study GWs from EMRIs in a static, spherically symmetric nonvacuum spacetime describing a MBH immersed in generic astrophysical environment was established in Ref. \cite{Cardoso:2022whc}, 
it was shown how the astrophysical environment changes GW emissions and the ability of GW to constrain smaller scale matter distribution around BHs.

The above discussions are based on quasi-circular orbits or eccentric cases without GW reaction. 
Non-negligible eccentricity in the detector band is possible through dynamical channels, field triples, GW capture and triples near supermassive BHs \cite{Kozai:1962zz,Heggie:1975tg,Antonini:2017ash,Hoang:2019kye,Wen:2002km,Miller:2002pg}, 
so the inclusion of eccentricity is particularly relevant for EMRIs 
although the eccentricity decreases due to GW radiation  \cite{Cutler:1994pb,Gair:2005ih,Cardoso:2020iji,Jiang:2021htl,Barsanti:2022ana}.
On the other hand, environmental effects are capable of increasing the eccentricity, 
even a small initial eccentricity can become larger over a long evolving time \cite{Yue:2019ozq,Cardoso:2020iji,Dai:2021olt}. 
The multiple harmonics induced by the eccentricity can break the degeneracy between source parameters which is helpful for the measurement of the position and distance of the GW source \cite{Yang:2022tig}.
This motivates us to study EMRIs in eccentric orbits.
The evolution of EMRIs within DM halos is not only affected by the GW reaction and the gravitational effect of DM halo,
but also influenced by the dynamical friction and accretion of the DM medium. 
When the SCO passes through the DM halo, the gravitational pull from the DM medium slows it down, this effect is called dynamical friction \cite{Chandrasekhar:1943ys, Cardoso:2020iji}.
If the SCO is a small BH, it will accrete the surrounding DM medium around it.
The accretion rate can be described by the Bondi-Hoyle-Lyttleton accretion model \cite{Bondi:1944jm,Edgar:2004mk}.

In this paper, we study eccentric orbital motions and GWs of EMRIs in galaxies with DM halos.
We discuss the combined effects of GW reaction, dynamical friction and accretion on the orbital motion of EMRIs.
The paper is organized as follows.
In Sec. \ref{motion}, a review of the spacetime of galaxies harboring MBHs is given first,
then we discuss the geodesic motion in this spacetime.
In Sec. \ref{gw-emris}, we use the "Numerical Kludge" method \cite{Gair:2005ih,Gair:2005is,Babak:2006uv} to calculate GWs from eccentric EMRIs in galaxies with DM halos.
To assess the capability of detecting DM halos with LISA, we calculate the mismatch between GWs from EMRIs with and without DM halos along with their signal-to-noise ratios (SNRs).
To consider the influence of the degeneracy between the source parameters, we use
the Fisher information matrix (FIM) method to estimate the parameter errors 
in Sec. \ref{sec:fim}.
We draw conclusions in Sec. \ref{conclusion}.
In this paper, we use the units $G=c=1$.

\section{The motions of binaries in the environments of galaxies}
\label{motion}

The density distribution of a galactic DM halo can be parameterized as \cite{Taylor:2002zd}
\begin{equation}
    \label{general-density}
    \rho(r)=2^{(\gamma-\alpha)/k} \rho_0 (r/r_0)^{-\alpha}(1+r^k / r_0^k)^{-(\gamma-\alpha)/k},
\end{equation}
where $r$ is the distance to the center of the halo, $r_0$ is the typical length-scale of a galaxy, $\rho_0$ is the density of the DM halo at $r_0$,
and the parameters $\alpha$, $\gamma$, and $k$ describe the type of DM halo depending on the size, mass, and form of the galaxy.
For the Navarro, Frenk and White profile \cite{Navarro:1996gj} which models galaxies with the largest content of DM, $\alpha=1$, $\gamma=3$ and $k=1$.
For a dwarf galaxy composed of about a thousand up to several billion stars, 
the Burkert model with $\alpha=1$, $\gamma=3$ and $k=2$ \cite{Burkert:1995yz} applies.
For the Hernquist-type density distribution \cite{Hernquist:1990be}
describing the S\'{e}rsic profiles observed in the bulges and elliptical galaxies,
$\alpha=1$, $\gamma=4$ and $k=1$.
In this paper, without the loss of generality, 
we focus on the Hernquist-type density distribution investigated in \cite{Cardoso:2021wlq}.
Note that the orbital motion and GW waveform will be different for different choices of the DM density profile \cite{Konoplya:2022hbl,Figueiredo:2023gas,Speeney:2024mas}.

The Hernquist-type density distribution can be rewritten as 
\begin{equation}
    \label{h-density}
    \rho_\text{H}=\frac{M r_0}{2\pi r (r+r_0)^3},
\end{equation}
where $M$ is the total mass of the DM halo.
For astrophysical scenarios, the compactness of the DM halo $M/r_0$ is usually small, i.e., $M/r_0\lesssim 10^{-4}$ \cite{Navarro:1996gj},
and $r_0=20$ kpc for the Milky Way \cite{Pierre:2014tra,Iocco:2015xga}.
In the galactic environments containing MBHs, the parameter $M/r_0$ is basically free \cite{Cardoso:2021wlq}.
Following Ref. \cite{Cardoso:2021wlq}, 
the energy-momentum tensor of a galaxy harboring a MBH with the mass $M_\text{BH}$ is assumed to be
\begin{equation}
\label{emt-halo}
    T^\mu_\nu={\rm diag}(-\rho_\text{DM},0,P_t,P_t),
\end{equation}
and the spacetime geometry is described by the static, spherically symmetric metric
\begin{equation}
    \label{h-metric}
    ds^2=-f(r) dt^2+\frac{dr^2}{1-2m(r)/r}+r^2(d\theta^2+\sin^2\theta\,d\phi^2).
\end{equation}
Combining Eqs. \eqref{emt-halo} and \eqref{h-metric}, Einstein equations become
\begin{gather}
\label{f-halo}
    \frac{1}{f(r)}\frac{d f(r)}{d r}=\frac{2m(r)}{r(r-2m(r))},\\
    \label{rhom-halo}
    \frac{dm(r)}{dr}=4\pi r^2\rho_\text{DM}(r),\\
    \label{pt-halo}
    2P_t=\frac{m(r)\rho_\text{DM}(r)}{r-2m(r)}.    
\end{gather}
Dependent on the choice of $\rho_\text{DM}(r)$, the mass function $m(r)$ will be different.
We assume that the  density profile for a MBH residing at the center of the Hernquist-type distribution \eqref{h-density} is
\begin{equation}
    \label{dm-density}
    \rho_\text{DM}=\frac{Mr_0(1+2M_{\rm BH}/r_0)(1-2M_{\rm BH}/r)}{2\pi r(r+r_0)^3}.
\end{equation}
Obviously, in the absence of the central MBH, 
the density profile \eqref{dm-density} reduces to Eq. \eqref{h-density}.
At large distances, $r\gg M_\text{BH}$, the density profile $\rho_\text{DM}$ becomes the Hernquist-type distribution \eqref{h-density} for large galaxies with $r_0\gg M_\text{BH}$.
If $r_0\gg r\gg M_\text{BH}$, then
$\rho_\text{DM}\sim (M/r_0)^2/(Mr)$, 
so the DM density $\rho_\text{DM}$ is smaller if the compactness $M/r_0$ is smaller with fixed $M$, or if $M$ is larger with fixed compactness $M/r_0$.

With the choice of the density profile \eqref{dm-density},
we get the mass function
\begin{equation}
    \label{mr}
    m(r)=M_{\rm BH}+\frac{M r^2}{(r_0+r)^2}\left(1-\frac{2M_{\rm BH}}{r}\right)^2.
\end{equation}
The mass function $m(r)$ in Eq. \eqref{mr} is a combination of the original mass distribution of Hernquist-halo and the mass of the central BH.

Combining Eqs. \eqref{f-halo} and \eqref{mr}, we get \cite{Cardoso:2021wlq}
\begin{equation}
    \label{f-e}
\begin{split}
    f(r)&=\left(1-\frac{2M_{\rm BH}}{r}\right)e^\Upsilon,\\
    \Upsilon&=-\pi\sqrt{\frac{M}{\xi}}+2\sqrt{\frac{M}{\xi}}\arctan\left(\frac{r+r_0-M}{\sqrt{M\xi}}\right),\\
    \xi&=2r_0-M+4M_{\rm BH}.
\end{split}
\end{equation}
The metric \eqref{h-metric} describes a BH spacetime with a horizon at $r=2M_{\rm BH}$ and a curvature singularity at $r=0$, the matter density vanishes at the horizon and the ADM mass of the spacetime is $M+M_\text{BH}$.
In the absence of DM halo, $M=0$, the spacetime \eqref{h-metric} reduces to  Schwarzschild BH with mass $M_\text{BH}$.
In general astrophysical environments, 
$M_\text{BH}\ll M \ll r_0$.
Expanding the function $f(r)$ in Eq. \eqref{f-e} about $M/r_0=0$ to the second order we get
\begin{equation}
    \label{f-a2}
    \begin{split}
    f(r)&\simeq \l(1-\frac{2M_{\rm BH}}{r}\r)
    \l(1-\frac{2M}{r_0}+\frac{4M^2}{3r_0^2}+\frac{2Mr}{r_0^2}\r)\\
    &=\l(1-\frac{2M_{\rm BH}}{r}\r)(1+\lambda+\beta r),
    \end{split}
\end{equation}
where $\lambda=-2M/r_0+4M^2/{3r_0^2}$ and $\beta=2M/{r_0^2}$.

Now, we consider a MBH in the center of a DM halo and a SCO moving on geodesics around the MBH in the equatorial plane ($\theta=\pi/2$).
The geodesic equation is
\begin{equation}
    \label{h-geodesic}
    \frac{d u_{\mu}}{d\tau}=\frac{1}{2} u^{\alpha} u^{\beta}\pa_{\mu}g_{\alpha\beta},
\end{equation}
where $u^{\alpha}=dr^{\alpha}/d\tau$, $\tau$ is the proper time, and $r^\alpha=(t,r,\theta,\phi)$.
Because the spacetime is static and spherically symmetric, from the geodesic equation \eqref{h-geodesic} we obtain two conserved quantities, $u_0=-E/\mu$ and $u_{\phi}=L/\mu$,
\begin{align}
    \label{u_0}
    u_0 &=-E/\mu=-\sqrt{1+2\varepsilon},\\
    \label{u_phi}
    u_{\phi}&=L/\mu=  h, 
\end{align}
where $E$ and $L$ represent the orbital energy and angular momentum of the system, respectively, and the reduced mass $\mu$ is approximately equal to the mass of the SCO.
The radial equation of motion is
\begin{equation}
    \label{eom}
    1+\left(\frac{dr}{d\tau}\right)^2\l(1-\frac{2m(r)}{r}\r)^{-1}+\frac{h^2}{r^2}=\frac{1+2\varepsilon}{f}.
\end{equation}

\begin{figure*}[htp]
    \centering
     \includegraphics[width=0.90\linewidth]{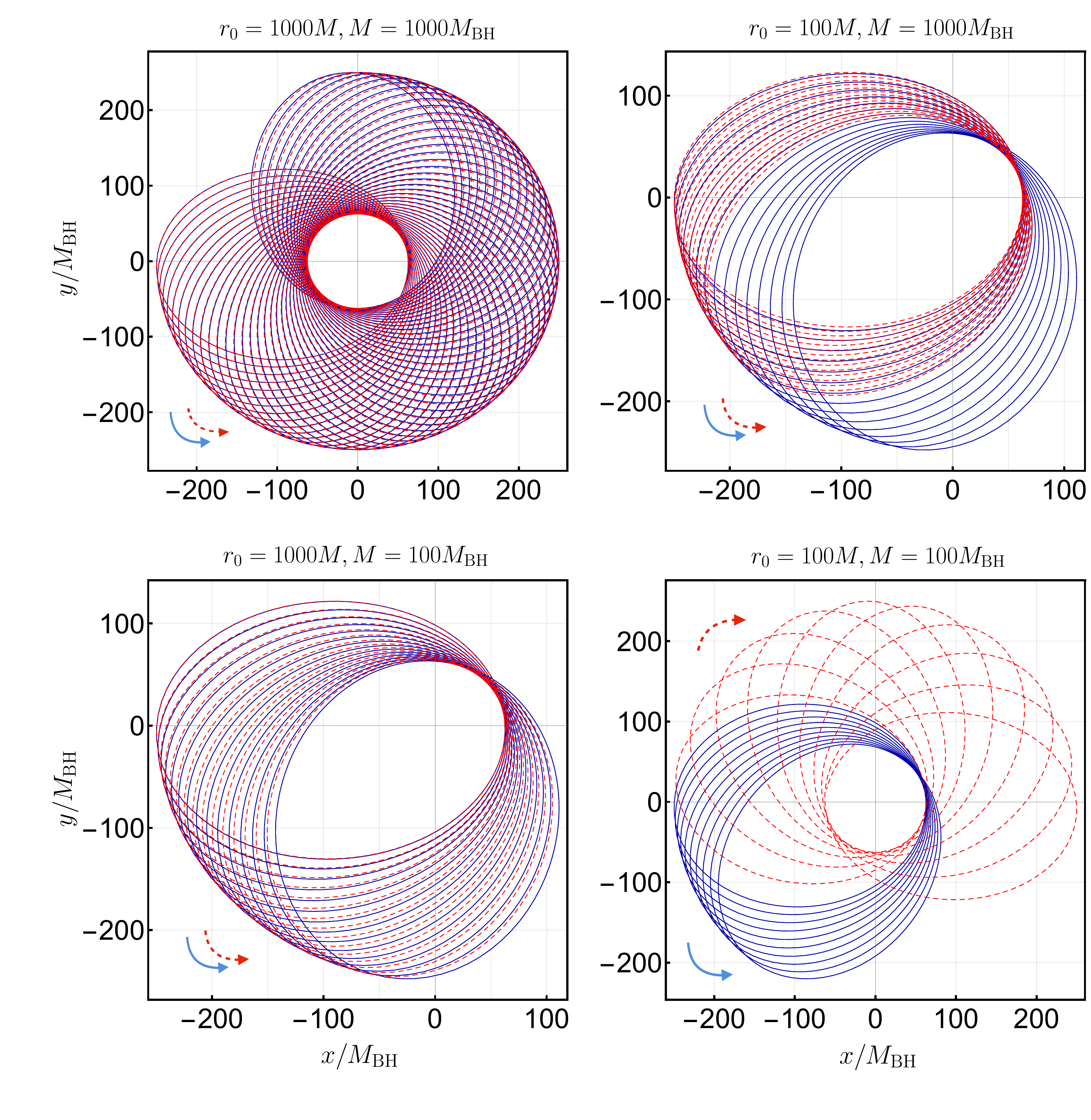}
    \caption{
   The orbits of EMRIs in galaxies with and without DM halos.
   The mass of the MBH is set as $M_{\text{BH}}=10^6 M_{\odot}$, the eccentricity $e=0.6$, and the semi-latus rectum $p=200 M_\text{BH}$.  
   We take the compactness $M/r_0$ as $10^{-2}$ and $10^{-3}$, and the 
   mass of the halo $M$ as $10^2 M_\text{BH}$ and $10^3 M_\text{BH}$.
   The red dashed lines show the trajectories with DM and 
   the blue solid lines show the orbits without DM.
   The arrows represent the directions of orbital precessions.
   }
     \label{fig:2.01}
\end{figure*}

For convenience, we introduce the orbital elements, 
the semi-latus rectum $p$ and the eccentricity $e$, to parameterize the orbital motion,
\begin{equation}
    \label{rpe}
    r=\frac{p}{1+e\cos\chi},
\end{equation}
where $\chi$ is a parameter.
Rewriting the variables $h$ and $\varepsilon$ in terms of $p$ and $e$, we obtain
\begin{equation}
    \begin{split}
    \label{h^2}
    h^2=&\frac{p\,R_s\,(1+\lambda)+p^3\beta\,(1-e^2)^{-1}}{\l(1-\frac{R_s}{2p}(3+e^2)\r)+p\,\beta\,\l(1-\frac{2R_s}{p}\r)}\\
    &\times\frac{1}{2(1+\lambda)},
    \end{split}
\end{equation}
\begin{equation}
     \label{energe}
    \varepsilon=-\frac{\frac{R_s}{2p}(1-e^2)\l(1-\frac{2R_s}{p}\r)+\lambda\Sigma+\lambda^2\Delta+\beta Z}{2\left[\l(1-\frac{1}{2}\frac{R_s}{p}(3+e^2)\r)(1+\lambda)+p\,\beta\l(\frac{1}{2}-\frac{R_s}{p}\r)\right]}
\end{equation}
where $R_s=2M_\text{BH}$,
\begin{equation}
    \begin{split}
    \Sigma&=-\l(1-\frac{2R_s}{p}\r)+\frac{R_s}{2p}\l(1-\frac{4R_s}{p}\r)(1-e^2)\n,\\
    \Delta&=-\l(1-\frac{2R_s}{p}\r)-\frac{R_s^2}{p^2}(1-e^2)\n,\\
    Z&=-\frac{p(3+e^2)}{2(1-e^2)}\l(1-2\frac{R_s}{p}\r)-\frac{2R_s^2}{p}\n.
    \end{split}
\end{equation}
In terms of $\chi$, Eqs. \eqref{u_0} and \eqref{u_phi} become
\begin{equation}
\label{dphidc}
\begin{split}    
    \frac{d\phi}{d\chi}&=\l[\frac{1}{2}\frac{R_s}{p}(1+\lambda)+\frac{1}{2}p\beta(1-e^2)^{-1} \r]^{\frac{1}{2}}\\
    &\quad \times\bigg\{\frac{1}{2}\frac{R_s}{p}\l[1-\frac{R_s}{p}\l(3+e\cos\chi\r)\r]\\
    &\quad +\lambda\,A+\frac{1}{2}\lambda^2\,A +\beta\,B\bigg\}^{-\frac{1}{2}}J_1,
\end{split}
\end{equation}
\begin{equation}
\begin{split}
    \label{dtdc}
    \frac{dt}{d\chi}=&\frac{p}{(1+e\cos\chi)^2}\bigg\{ \l[1-(1+e)\frac{R_s}{p}\r]\\
    &\times\l[1-(1-e)\frac{R_s}{p}\r]+ C \bigg\}^{\frac{1}{2}}\\
    &\times\l[1-\frac{R_s}{p}(1+e\cos\chi) \r]^{-1}\\
    &\times\bigg\{ \frac{1}{2}\frac{R_s}{p}\l[1-\frac{R_s}{p}(3+e\cos\chi)\r]\\
    &+\lambda A+2\lambda^2 A+\beta B \bigg\}^{-\frac{1}{2}}J_2,
\end{split}
\end{equation}
where 
\begin{equation}
    \begin{split}
    A&=\frac{R_s}{p}\l[1-\frac{R_s}{p}(3+e\cos\chi)\r]\n,\\
    B&=\frac{p}{2(1-e^2)(1+e\cos\chi)}\biggl\{ 2\l(1-\frac{R_s}{p}\r)+\n\\
    &\l[1-\frac{4R_s}{p}-\l(\frac{R_s}{p}\r)^2(1-e^2)\r](1+e\cos\chi)\\
    &-\frac{R_s}{p}(1+\cos^2\chi)e^2 \biggr\}\n,\\
    C&=\lambda\l[1-\frac{1}{2}(3+e^2)\frac{R_s}{p}\r]+\frac{1}{2}p\beta\l(1-2\frac{R_s}{p}\r)\\
    &-\l(\lambda \Sigma+\lambda^2 \Delta +\beta Z 
 \r)\n,\\
    \end{split}
\end{equation}
\begin{equation}
    \begin{split}
    J_1=&\l(1+\lambda+\frac{\beta p}{1+e\cos\chi}\r)^{\frac{1}{2}}\bigg\{ 1-\frac{2M p/(1+e\cos\chi)}{\l[r_0+p/(1+e\cos\chi)\r]^2}\\
    &\times\l[ 1-\frac{R_s}{p}(1+e\cos\chi) \r] \bigg\}^{-\frac{1}{2}}\n,\\
    J_2=&\l(1+\lambda+\frac{\beta p}{1+e\cos\chi}\r)^{-\frac{1}{2}}\bigg\{ 1
    -\frac{2M p/(1+e\cos\chi)}{\l[r_0+p/(1+e\cos\chi)\r]^2}\\
    &\times\l[ 1-\frac{R_s}{p}(1+e\cos\chi) \r] \bigg\}^{-\frac{1}{2}}.
    \end{split}
\end{equation}
Equations. \eqref{dphidc} and \eqref{dtdc} can be integrated to obtain $\phi(\chi)$ and $t(\chi)$.
Taking different compactness and mass for the DM halo,
using Cartesian coordinates $(x, y)=(r\cos{\phi}, r\sin{\phi})$ in the equatorial plane,
we show the orbits of EMRIs in galaxies with and without DM in Fig. \ref{fig:2.01}.
Due to the gravitational force of DM halos, the orbits with DM halos are different from those without DM.
From Fig. \ref{fig:2.01}, we see that for the same value of $M$, 
the effect of DM halos on the orbital precession is larger if the compactness of the DM halo $M/r_0$ is bigger.
DM halos decrease the orbital precession rates, 
and they can even reverse the direction of precession if the density of the DM halo $\rho_\text{DM}$ is large enough.
The result of retrograde precessions of the orbital motion in the spacetime \eqref{h-metric} is consistent with that found in \cite{Destounis:2022obl}, and the anomalous precessions of binaries in DM environments were also found in \cite{Dai:2021olt,Igata:2022nkt,Igata:2022rcm}.

To probe DM halos and study their impact on the orbits of EMRIs, we calculate the time $T$ and the orbital precession $\Delta\phi$ over one cycle when the orbital parameter $\chi$ increases by $2\pi$,
\begin{align}
    \label{period}
    T&=\int_{0}^{2\pi}\frac{dt}{d\chi}d\chi,\\
    \label{pressesion}
    \Delta\phi&=\int_{0}^{2\pi}\frac{d\phi}{d\chi}d\chi-2\pi.
\end{align}
Expanding Eqs. \eqref{dphidc} and \eqref{dtdc} about $R_s/p=0$ to the second order and substituting the results into Eqs. \eqref{period} and \eqref{pressesion}, we get
\begin{equation}
    \begin{split}
        \label{a-period}
    T=&2\pi\sqrt{\frac{2p^3}{R_s}}\frac{1}{(1-e^2)^{3/2}}\bigg\{ 1+\frac{3}{2}(1-e^2)\frac{R_s}{p}\\
    &+\frac{3}{2}(1-e^2)\l[1+\frac{5}{4}(1-e^2)^{\frac{1}{2}}\r]\l( \frac{R_s}{p} \r)^2\\
    &+\frac{M}{r_0}
    +\frac{3(1-e^2)}{2}\frac{M R_s}{r_0 p}+\frac{5M^2}{6r_0^2}\\
    &+\frac{M p}{r_0^2(1-e^2)}\l(e^2-\frac{11}{2}\r)-\frac{3M p^2/{R_s}}{r_0^2(1-e^2)^2} \bigg\},\\
    \end{split}
\end{equation}
\begin{equation}
    \begin{split}
         \label{a-pression}
    \Delta\phi=&3\pi\frac{R_s}{p}+\frac{3\pi}{8}(18+e^2)\l(\frac{R_s}{p}\r)^2\\
    &-\frac{6\pi}{1-e^2}\frac{M p}{r_0^2}\l[ 1+\frac{1+e^2+2{p}/{R_s}}{3(1-e^2)^{1/2}} \r].
    \end{split}
\end{equation}

\begin{figure*}[htp]
    \centering
    \includegraphics[width=0.96\linewidth]{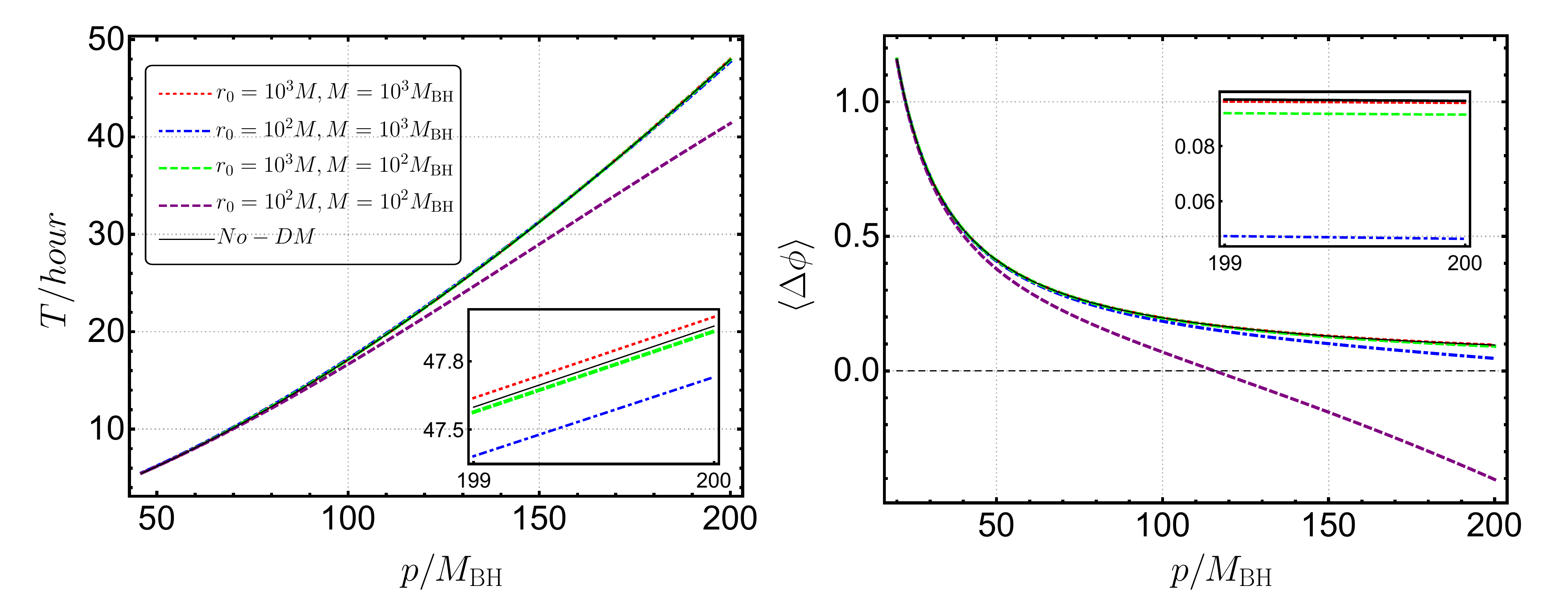}
    \caption{
    The results of the orbital period and precession for EMRIs in galaxies with and without DM.
    The mass of the central MBH is set as $M_{\text{BH}}=10^6 M_{\odot}$ and the eccentricity $e=0.6$.  
    We take the compactness $M/r_0$ as $10^{-2}$ and $10^{-3}$, and the total mass $M$ of the halo as $10^3 M_\text{BH}$, $10^2 M_\text{BH}$ and $M=0$. The inserts show the evolution in a short time period.
    }
    \label{fig:2.02}
\end{figure*}

The terms with $M$ in the above Eqs. \eqref{a-period} and \eqref{a-pression} come from DM halos.
In the absence of DM, $M=0$, the above results \eqref{a-period} and \eqref{a-pression} recover those for EMRIs with the central MBH being a Schwarzschild BH.
The dominant contribution to the period $T$ in Eq. \eqref{a-period} is the first term, 
so $T$ becomes larger as the semi-latus rectum $p$ increases. 
However, there are both positive and negative contributions to $T$ from the local DM halos,  
so the local DM halos may slow down the increase of $T$ as $p$ increases because of the negative contribution in the last term in Eq. \eqref{a-period}
and the presence of DM halos helps the increase of $T$ with $p$ if the last negative contribution is negligible.
From Eq. \eqref{a-pression}, it is easy to understand that the presence of DM halo decreases the orbital precession
and even retrogrades the orbital precession if the local density of DM halos $\rho_\text{DM}\sim M/r_0^2$ is sufficiently large that the third term dominates over the first two terms.
For larger orbits, i.e., the semi-latus rectum $p$ is larger, 
the orbital precession decreases and the prograde precession decreases faster in the presence of DM halos because the third term due to DM halos in Eq. \eqref{a-pression} becomes bigger.
In the presence of DM halos, the prograde-to-retrograde precession transition happens at some critical value of $p$, and then the retrograde precessions become bigger if $p$ is larger. 
Choosing different values for the compactness $M/r_0$ and the total mass of DM halos $M$ and using Eqs. \eqref{a-period} and \eqref{a-pression}, 
we plot the results of the period $T$ and the orbital precession $\Delta\phi$ versus the semi-latus rectum $p$ in Fig. \ref{fig:2.02}.
As expected, the orbital period $T$ increases with $p$, the prograde precessions decrease with $p$
and DM halos accelerate the decrease.
For the case of $r_0=10^{2}M$ and $M=10^2 M_\text{BH}$, the periapsis shifts change from prograde precessions to retrograde precessions at $p= 120 M_\text{BH}$ and the retrograde precession increases with $p$ when $p\gtrsim 120 M_\text{BH}$.

From the above discussions, we see that the orbital motions of EMRIs are influenced by DM halos,
and we expect that the effects of local DM halos will leave imprints on GWs
so that we can probe local DM halos through the observations of GWs emitted from EMRIs.

\section{GWs of EMRIs in the environments of galaxies}
\label{gw-emris}

\subsection{Orbital evolution}
\label{orbit-evolve}

In this section, we take the SCO as a BH.
When the small BH passes through the DM medium, the gravitational pull from DM particles decelerates the small BH. 
This effect is called the dynamical friction
and the force is
\begin{equation}
\label{fDF}
    \bm{f}_\text{DF}=-\frac{4\pi \mu^2 \rho_\text{DM} \ln\Lambda}{v^3}\bm{v},
\end{equation}
where $\bm{v}$ is the velocity of the small BH, 
and we choose the Coulomb logarithm $\ln\Lambda=3$ \cite{Eda:2014kra}.

Combining Eqs. \eqref{rpe}, \eqref{dphidc}, \eqref{dtdc}, and \eqref{fDF}, the energy and the angular momentum loss rates caused by the dynamic frictions are
\begin{equation}
\begin{split}
\label{dEdtDF}
    \left(\frac{dE}{dt}\right)_{\text{DF}}=&-\frac{6\sqrt{2}\mu^2(1-R_s/p)^{3/2}}{\sqrt{R_s}(p-(1+e\cos\chi) R_s)}\frac{M}{r_0^2}K_1\\
    &\times\sqrt{(p-(1-e)R_s)(p-(1+e)R_s)}\,,
\end{split}
\end{equation}
\begin{equation}
\begin{split}
\label{dLdtDF}
    \left(\frac{dL}{dt}\right)_{\text{DF}}=&-\frac{12\mu^2\sqrt{p}(p-R_s)^{5/2}}{R_s(p-R_s-eR_s\cos\chi)^{5/2}}\frac{M}{r_0^2}K_2\\
    &\times(p+(-1+e)R_s)(p-(1+e)R_s),
\end{split}
\end{equation}
where
\begin{equation}
\begin{split}
    K_1=&\bigg[(p-R_s)(1+e\cos\chi)^4\n\\
    &+e^2(p-3R_s-eR_s\cos\chi)\sin^2\chi\bigg]^{-1/2},\n\\
    K_2=&\bigg[(p-Rs)(1+e\cos\chi)^4\n\\
    &+e^2(p-3R_s-eR_s\cos\chi)\sin^2\chi\bigg]^{-3/2}.\n
\end{split}
\end{equation}

When moving in the DM halo, the small BH can accrete the surrounding DM medium. 
This effect can be modeled by the Bondi-Hoyle accretion \cite{Bondi:1944jm,Edgar:2004mk,Macedo:2013qea}
\begin{equation}
\label{dmdt}
    \dot{\mu} = \frac{4 \pi \rho_\text{DM} \mu^2}{(v^2+c_\text{s}^2)^{3/2}},
\end{equation}
where $c_\text{s}=\sqrt{\delta P_t/\delta \rho}$ is the sound speed, and the over-dot indicates differentiation with respect to $t$.
Note that the effects of DF and accretion in Eqs. \eqref{fDF} and \eqref{dmdt} are Newtonian estimated.

It was shown in Refs. \cite{Hughes:2018qxz,Blachier:2023ygh} that the accretion does not result in the exchange of orbital angular momentum, i.e., 
$\left({dL}/{dt}\right)_\text{acc}=0$.
The variation of orbital energy due to the mass increase is
\begin{equation}
\label{dEdtacc}
    \left(\frac{dE}{dt}\right)_\text{acc}
    =\frac{\dot{\mu}}{\mu}E-\dot{\mu}\, v^2,
\end{equation}
where $E$ is the orbital energy defined in Eq. \eqref{u_0}.

The energy and angular momentum loss rates due to GW emissions are described by the quadrupole formula
\begin{equation}
\label{dEdtGW1}
    \left(\frac{dE}{dt}\right)_{\text{GW}}=-\frac{1}{5}\dddot{\mathcal{I}}^{jk}\dddot{\mathcal{I}}^{jk},
\end{equation}
\begin{equation}
\label{dLdtGW1}
    \left(\frac{dL_i}{dt}\right)_{\text{GW}}=-\frac{2}{5}\epsilon_{ijk}\ddot{\mathcal{I}}^{jl}\dddot{\mathcal{I}}^{kl},
\end{equation}
where 
$\mathcal{I}^{jk}$ is the symmetric-traceless part of the mass quadrupole.
Using the results for the orbital motions of EMRIs within DM halos obtained above and considering the variation of the mass, 
we get the leading order energy and angular momentum fluxes of the GW reaction
\begin{equation}
\begin{split}
\label{dEdtGW}
    \left(\frac{dE}{dt}\right)_{\text{GW}}=&-\frac{4M_\text{BH}^5\mu^2}{15p^5}f_1(e,\phi)\left(1-\frac{6M}{r_0}\right)\\
    &+\frac{4 M_\text{BH}^{9/2}\dot{\mu}\mu}{5p^{7/2}}f_2(e,\phi)\left(1-\frac{5M}{r_0}\right)\\
    &+\frac{6M_\text{BH}^4\dot{\mu}^2}{5p^2}f_3(e,\phi)\left(1-\frac{4M}{r_0}\right),
\end{split}
\end{equation}
\begin{equation}
\begin{split}
\label{dLdtGW}
    \left(\frac{dL}{dt}\right)_{\text{GW}}=&-\frac{4M_\text{BH}^{9/2}\mu^2}{5p^{7/2}}g_1(e,\phi)\left(1-\frac{5M}{r_0}\right)\\
    &-\frac{24M_\text{BH}^4\mu\dot{\mu}}{5p^2}g_2(e,\phi)\left(1-\frac{4M}{r_0}\right)\\
    &+\frac{24M_\text{BH}^{7/2}\dot{\mu}^2}{5p^{1/2}}g_3(e,\phi)\left(1-\frac{3M}{r_0}\right),
\end{split}
\end{equation}
where
\begin{equation}
\begin{split}
f_1(e,\phi)=&\l(1+e\cos\phi\r)^4(24+13e^2 \n\\
&+48e\cos\phi+11e^2\cos2\phi),\n\\
f_2(e,\phi)=&(1+e\cos\phi)^2(18+13e^2+40e\cos\phi \n\\
&+9e^2\cos2\phi), \n\\
f_3(e,\phi)=&12+20e^2+4e^4+(36e+17e^3)\cos\phi \n\\
&+20e^2\cos2\phi+3e^3\cos3\phi,\n
\end{split}
\end{equation}
and
\begin{equation}
\begin{split}
    g_1(e,\phi)=&(1+e\cos\phi)^3(8+e^2\n\\
    &+12e\cos\phi+3e^2\cos2\phi),\n\\
    g_2(e,\phi)=&(1+e\cos\phi)^2\sin\phi,\n\\    
    g_3(e,\phi)=&2+e^2+3e\cos\phi.\n
\end{split}
\end{equation}
In Eqs. \eqref{dEdtGW} and \eqref{dLdtGW}, 
the factors $(1-n M/r_0)$ are the corrections from DM halos around the central MBH, 
and the terms with $\dot{\mu}$ are the contributions from the mass variation of the small BH.
Note that the losses of energy and angular momentum due to the GW reaction depend on the compactness $M/r_0$, 
and the energy flux becomes smaller if the compactness is larger.
In the absence of local DM halos, $M=0$, 
Eqs. \eqref{dEdtGW} and \eqref{dLdtGW} reduce to the results for eccentric binaries in the vacuum \cite{Peters:1963ux,Peters:1964zz}.

\begin{figure*}[htp]
    \centering
    \includegraphics[width=0.96\linewidth]{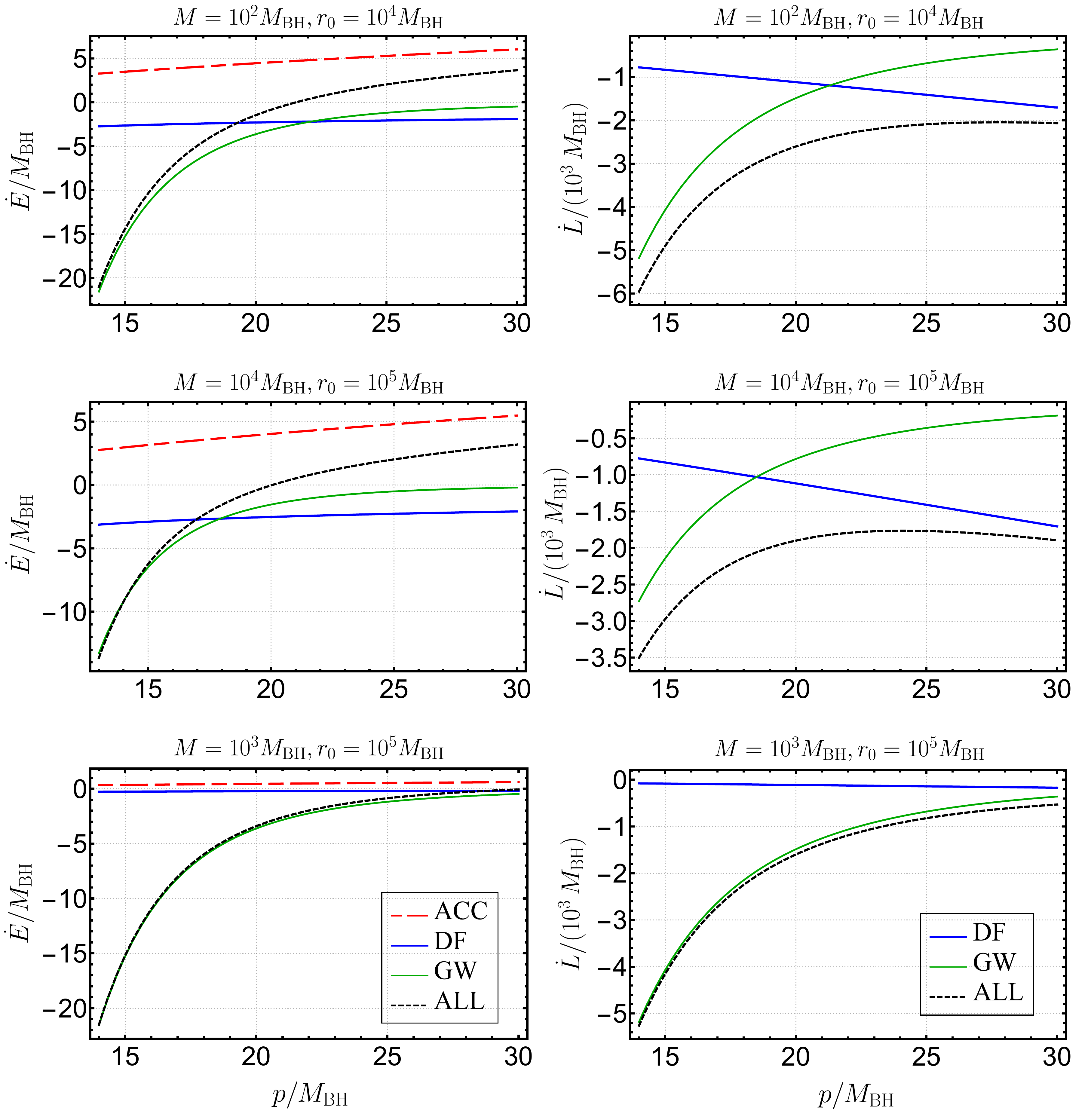}
    \caption{
    The averaged energy and angular momentum fluxes from the dynamical friction, accretion effect and GW reaction at different $p$.
    The symbols "ACC", "DF", and "GW" represent the accretion, dynamical friction and GW reaction, respectively.
    "ALL" means the combined net effect of the dynamical friction, accretion and GW reaction.
    The mass of the central MBH is chosen as $M_\text{BH}=10^6M_\odot$, the mass of the small BH is $\mu=10M_\odot$ and 
    the eccentricity is chosen as $e=0.1$.
    }
\label{fig:3.01EL}
\end{figure*}

The energy and angular momentum fluxes with the orbital period averaging are 
\begin{equation}
     \label{dEdtav3}
    \left<\frac{dE}{dt}\right>=\frac{1}{T}\int_{0}^{T} \frac{dE}{dt} \,dt=\frac{1}{T}\int_{0}^{2\pi} \frac{dE}{dt} \frac{dt}{d\phi}\,d\phi\,,
\end{equation}
\begin{equation}
    \label{dLdtav3}
    \left<\frac{dL_z}{dt}\right>=\frac{1}{T}\int_{0}^{T}
    \frac{dL_z}{dt} \,dt=\frac{1}{T}\int_{0}^{2\pi} \frac{dL_z}{dt} \frac{dt}{d\phi}\,d\phi\,.
\end{equation}
Inserting Eqs. \eqref{dEdtDF}, \eqref{dLdtDF}, \eqref{dEdtacc}, \eqref{dEdtGW}, and \eqref{dLdtGW} into Eqs. \eqref{dEdtav3} and \eqref{dLdtav3}, we obtain the averaged energy and angular momentum fluxes of the dynamical friction, accretion, and GW reaction.
We show the averaged energy and angular momentum fluxes of the dynamical friction, accretion, and GW reaction with different parameters for DM halos in Fig. \ref{fig:3.01EL}.

In the left panels of Fig. \ref{fig:3.01EL}, we graph the energy fluxes at different $p$ for different $M$ and $r_0$, and the angular momentum fluxes are shown in the right panels.
As expected from Eq. \eqref{dEdtacc}, the energy fluxes of the accretion are positive.
The energy fluxes of the GW reaction dominate when the small BH is close to the central MBH ($p \lesssim 15M_\text{BH}$), 
and the energy fluxes of the dynamical friction dominate when $p \gtrsim 25M_\text{BH}$. 
We also see that the angular momentum fluxes of the GW reaction are much bigger than the dynamical friction when $p \lesssim 15M_\text{BH}$, 
and the angular momentum fluxes of the dynamical friction are bigger than the GW reaction when $p \gtrsim 25M_\text{BH}$.  
From Eqs. \eqref{dEdtGW} and \eqref{dLdtGW}, 
we see that the fluxes of the GW reaction are inversely proportional to $p^\delta$, 
so the fluxes of the GW reaction are larger when the orbit of the binary is smaller.
From Eqs. \eqref{fDF} and \eqref{dmdt}, 
we see that $\dot{\mu}$ and the dynamical friction are proportional to $v^{-2}$. 
Since the velocity is smaller in larger orbits, 
thus the effects of the accretion and dynamical friction are larger when $p\gtrsim25M_\text{BH}$ and smaller when $p\lesssim15M_\text{BH}$.

The energy and angular momentum fluxes of the GW reaction shown in the upper and lower panels in Fig. \ref{fig:3.01EL} are almost the same, 
and the effects of the accretion and dynamical frictional shown in the upper and middle panels in Fig. \ref{fig:3.01EL} are almost the same.
This is because the compactness of the DM halos $M/r_0$ chosen in the upper and lower panels is the same, and the effect of DM halo on the energy and angular momentum fluxes of the GW reaction is mainly manifested through $M/r_0$.
since the same values of $M/r^2_0$ are chosen in the upper and middle panels, 
andthe effects of the dynamical friction and accretion are mainly related to the DM density $\rho_\text{DM}\sim M/r^2_0$, the energy fluxes due to the dynamical friction and accretion shown in the upper and middle panels are almost the same.

\begin{figure*}[htp]
    \centering
    \includegraphics[width=0.94\linewidth]{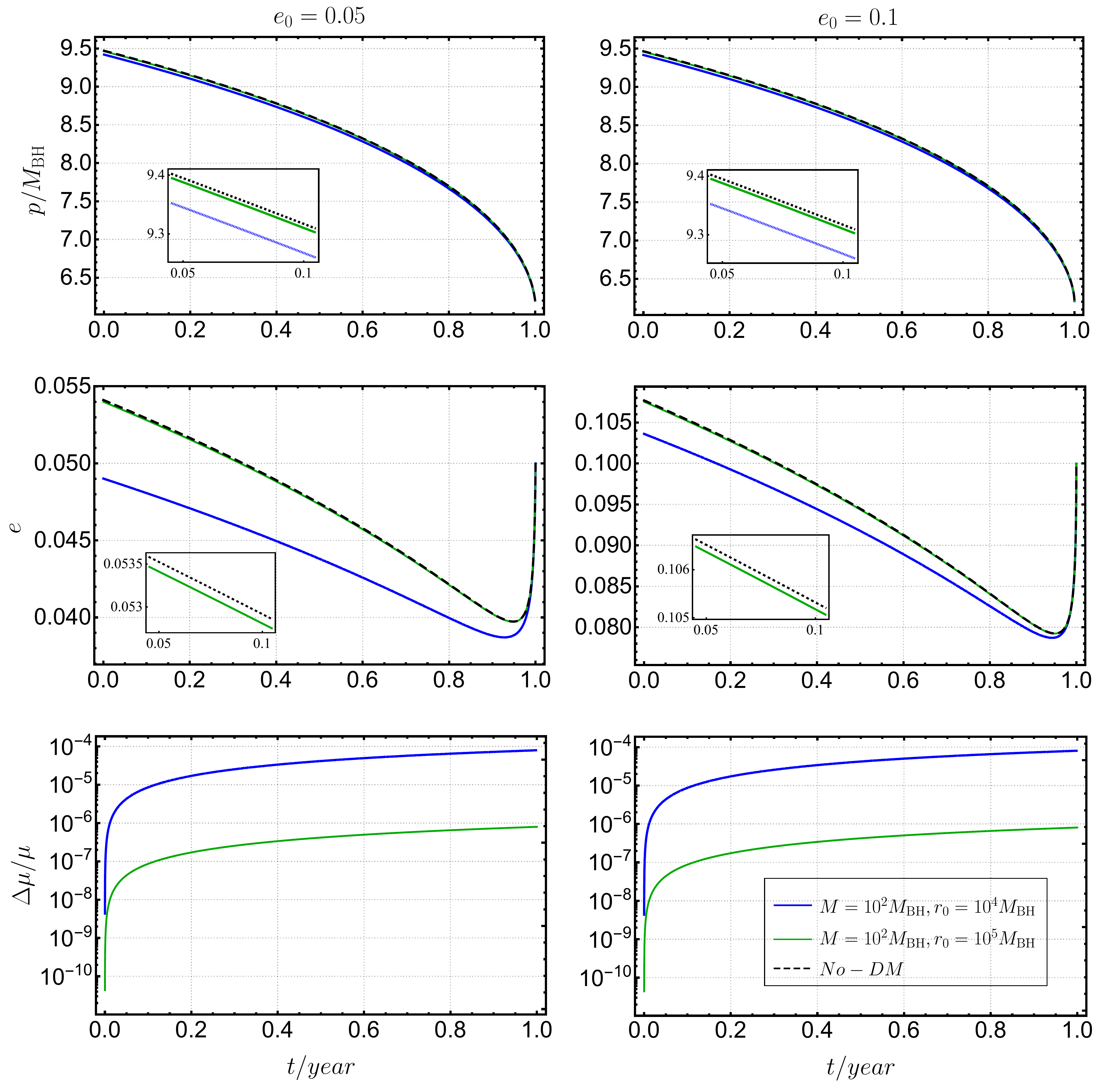}
    \caption{
    The evolution of the orbital parameters $p(t)$, $e(t)$, and $\mu(t)$.
    We choose the orbital evolution over one year before $p_0=6.2M_\text{BH}$, with initial eccentricity values of $e_0=0.05$ and $0.1$.
    The mass of the central MBH is chosen as $M_\text{BH}=10^6M_\odot$, the mass of the SCO is $\mu_0=10M_\odot$ and 
    we consider two different values for the compactness of the DM halos, $M/r_0=10^{-2}$ and $10^{-3}$. 
    The black dashed lines correspond to the cases without DM.
    }
\label{fig:3.01peu}
\end{figure*}

The changes of orbital energy and angular momentum due to the net effect of the dynamical friction, accretion and GW reaction can be written as
\begin{equation}
    \begin{split}
    \label{dE-B}
    \l(\frac{dE}{dt}\r)_{\text{orb}}=&\frac{dE}{dp}\frac{dp}{dt}+\frac{dE}{de}\frac{de}{dt}+\frac{dE}{d\mu}\frac{d\mu}{dt}\\
        =&\l<\frac{dE}{dt}\r>_{\text{GW}}+\l<\frac{dE}{dt}\r>_{\text{DF}}\\
        &+\l<\frac{dE}{dt}\r>_{\text{ACC}},
        \end{split}     
\end{equation}
\begin{equation}
    \begin{split}
    \label{dL-B}
    \l(\frac{dL}{dt}\r)_{\text{orb}}&=\frac{dL}{dp}\frac{dp}{dt}+\frac{dL}{de}\frac{de}{dt}+\frac{dL}{d\mu}\frac{d\mu}{dt}\\
    &=\l<\frac{dL}{dt}\r>_{\text{GW}}+\l<\frac{dL}{dt}\r>_{\text{DF}}.
    \end{split}     
\end{equation}
By combining the exact geodesic motion with the energy and angular momentum fluxes of DF, accretion, and GWs, we can obtain the orbital evolution of EMRIs in DM halos. Compared to the Teukolsky approach, the method we use here is less accurate but faster to calculate and more precise than the Newtonian method \cite{Gair:2005is}.
Inserting Eqs. \eqref{u_0}, \eqref{u_phi}, \eqref{dEdtDF}, \eqref{dLdtDF}, \eqref{dEdtacc}, \eqref{dEdtGW} and \eqref{dLdtGW} into Eqs. \eqref{dE-B} and \eqref{dL-B}, 
and numerically calculating the evolution of the orbital parameters $p$, $e$ and $\mu$ due to the combined effects of the dynamical friction, accretion, and GW reaction, 
we get the results as shown in Fig. \ref{fig:3.01peu}.
From Fig. \ref{fig:3.01peu} we see that EMRIs immersed in DM halos evolve more slowly.
The larger the compactness $M/r_0$ is, the more evolution time it takes.
The presence of DM halos also slows down the decrease rate of the eccentricity. 
Near $p_0=6.2M_\text{BH}$, the orbital eccentricity of EMRIs increases sharply, 
which is a significant feature of eccentric EMRIs \cite{Gair:2005is}.
In the one-year evolution, the mass of the small BH increases very little
and the mass increases less with smaller $\rho_\text{DM}$.

\subsection{Gravitational waveforms analysis}
\label{gw-analysis}

\begin{figure*}[htbp]
   \includegraphics[width=.94\textwidth,origin=c]{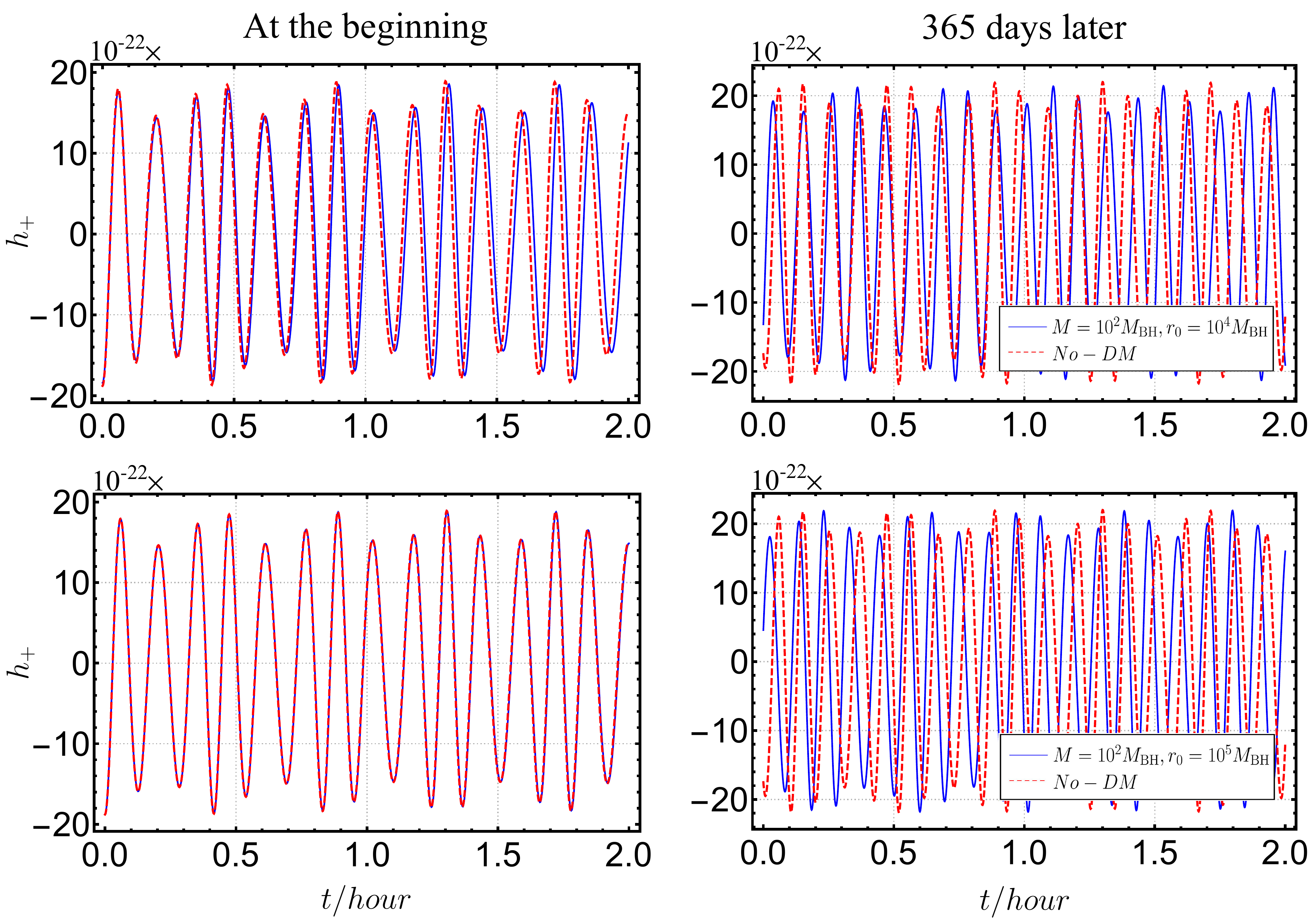}
   \caption{
   The time-domain plus mode GW waveforms for EMRIs with and without DM halos.
   The mass of the central MBH is $M_\text{BH}=10^6M_\odot$, the mass of the SCO is $\mu=10M_\odot$, the total mass of DM halos $M=10^2M_\text{BH}$, 
   the inclination angle $\iota=\pi/6$, the luminosity distance $d_L=1$ Gpc,
   the initial longitude of the pericenter $\omega_0=0$, and the initial eccentricity $e_0=0.1$ at $p_0=10M_\text{BH}$. $M=0$ corresponds to the case without DM halos.
   The left panels show the initial waveforms.
   The right panels show the waveforms after one year.
   The top panels are for $M/r_0=10^{-2}$, and the bottom panels are for $M/r_0=10^{-3}$. 
   }
   \label{fig:3.03}
\end{figure*}

As discussed above, the effects of DM halos will be manifested in GW waveforms.
The quadrupole formula of GWs is
\begin{equation}
\label{h-jk}
h^{jk}=\frac{2}{d_L} {\ddot{I}}^{jk},
\end{equation}
where $d_L$ is the luminosity distance between the detector and the source,
and $I_{jk}$ is the quadrupole moment of EMRIs.
The plus and cross tensor modes $h_+$ and $h_{\times }$ in the transverse-traceless gauge are given by 
\begin{align}
    \label{hplus}
    h_+&=\frac{1}{2}\l(e^j_X e^k_X-e^j_Y e^k_Y\r)h_{jk},\\
    \label{hcross}
    h_{\times}&=\frac{1}{2}\l(e^j_X e^k_Y+e^j_Y e^k_X\r)h_{jk},
\end{align}
where $e_X$ and $e_Y$ are the orthonormal vectors in the plane that are perpendicular to the direction from the detector to the GW source.
Plugging the results for the orbital evolution obtained above into Eq. \eqref{h-jk}, we numerically calculate the time-domain GW waveforms.
The time-domain plus-mode GW waveforms for EMRIs with and without DM halos are shown in Fig. \ref{fig:3.03}.
From Fig \ref{fig:3.03}, we see that initially the difference between GW waveforms with and without DM halos is negligible.
One year later, the two waveforms for EMRIs with and without DM halos are quite different.

\begin{figure}[htbp]
   \includegraphics[width=.46\textwidth,origin=c]{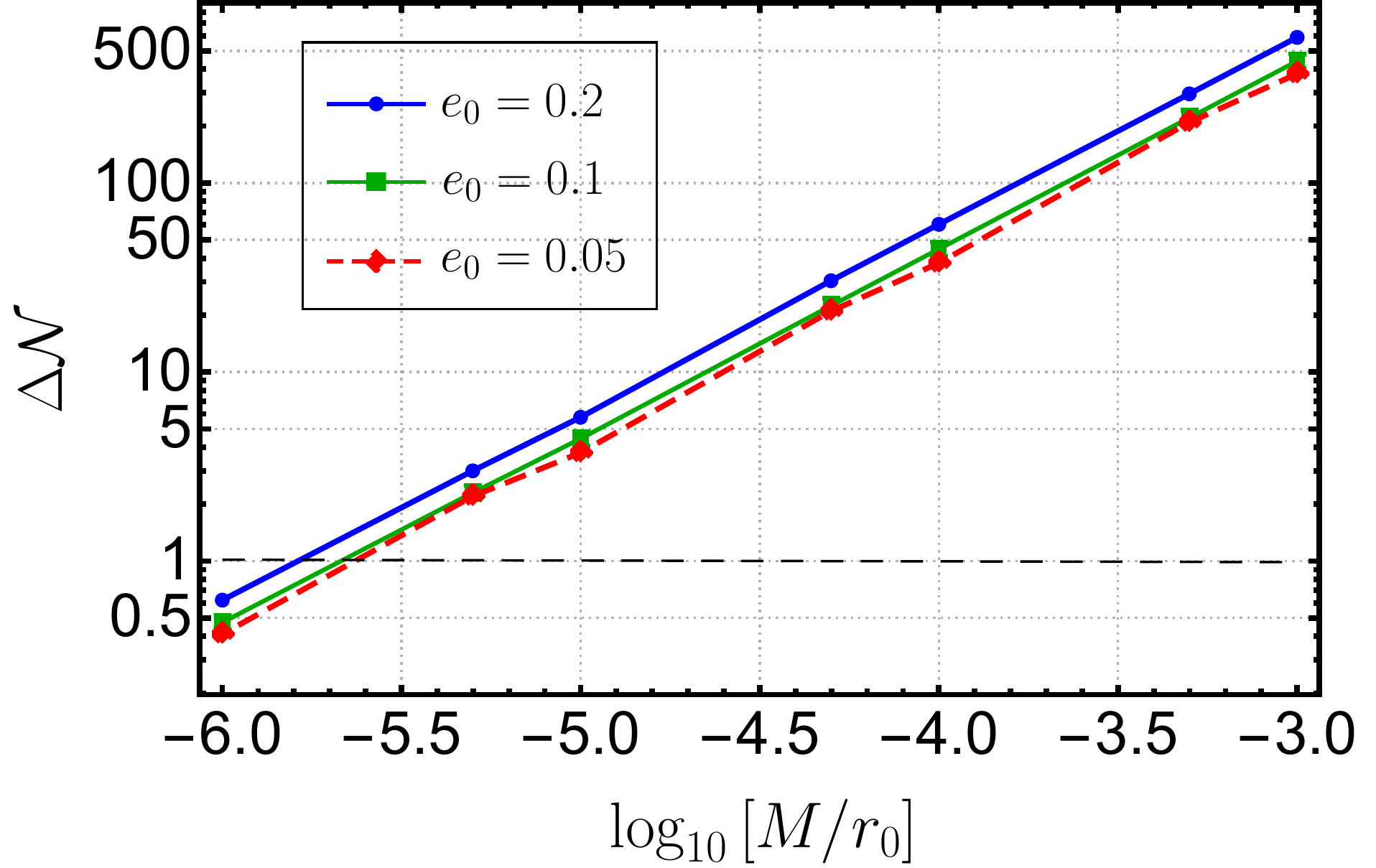}
   \caption{
   The difference between the orbital cycles with and without DM halos $\Delta\mathcal{N}(t)$ accumulated over one-year evolution for different compactness values of DM halos.
   The mass of the central MBH is $M_\text{BH}=10^6M_\odot$, the initial mass of the SCO is $\mu_0=10M_\odot$ 
   and the mass of DM halos is $M=10^2 M_\text{BH}$.
   The initial eccentricity values are chosen as $0.05$, $0.1$ and $0.2$.
   We take the initial semi-latus rectum $p_0$ at the position where the orbital period $T_0=2\pi\sqrt{{(5R_s)^3}/{M_\text{BH}}}$.
   The black dashed line corresponds to $\Delta\mathcal{N}=1\,\text{rad}$.
   }
   \label{fig:3.04}
\end{figure}

To quantify the impact of DM halo environments on the dephasing of GW waveforms, 
we calculate the number of orbital cycles accumulated from initial time $t_i$ to the final time $t_f$ \cite{Berti:2004bd, Kavanagh:2020cfn,Barsanti:2022ana}
\begin{equation}\label{n-gw}
   \mathcal{N}(t)=\int_{t_i}^{t_f} \dot\phi(t) dt.
\end{equation}
Over one-year evolution ($t_f-t_i=1$ year) starting from the time $t_i$ when the orbital period $T_0=2\pi\sqrt{{(5R_s)^3}/{M_\text{BH}}}$, 
the numbers of orbital cycles accumulated for EMRIs with and without DM halos are $\mathcal{N}_\text{DM}$ and $\mathcal{N}_0$, respectively.
In Fig. \ref{fig:3.04}, we show the difference between the number of orbital cycles with and without DM halos accumulated over one year, $\Delta\mathcal{N}=\mathcal{N}_\text{DM}-\mathcal{N}_\text{0}$.
Following \cite{Maselli:2020zgv}, we choose $\Delta\mathcal{N}\sim 1\,\text{rad}$ as the threshold for a detectable dephasing.
Note that this threshold cannot be taken as a sufficient condition for detectability.
The results show that the compactness can be detected as small as $M/r_0\lesssim 10^{-5}$, 
and eccentric orbits can detect smaller compactness than circular orbits.

\begin{figure}[htbp]
   \includegraphics[width=.47\textwidth,origin=c]{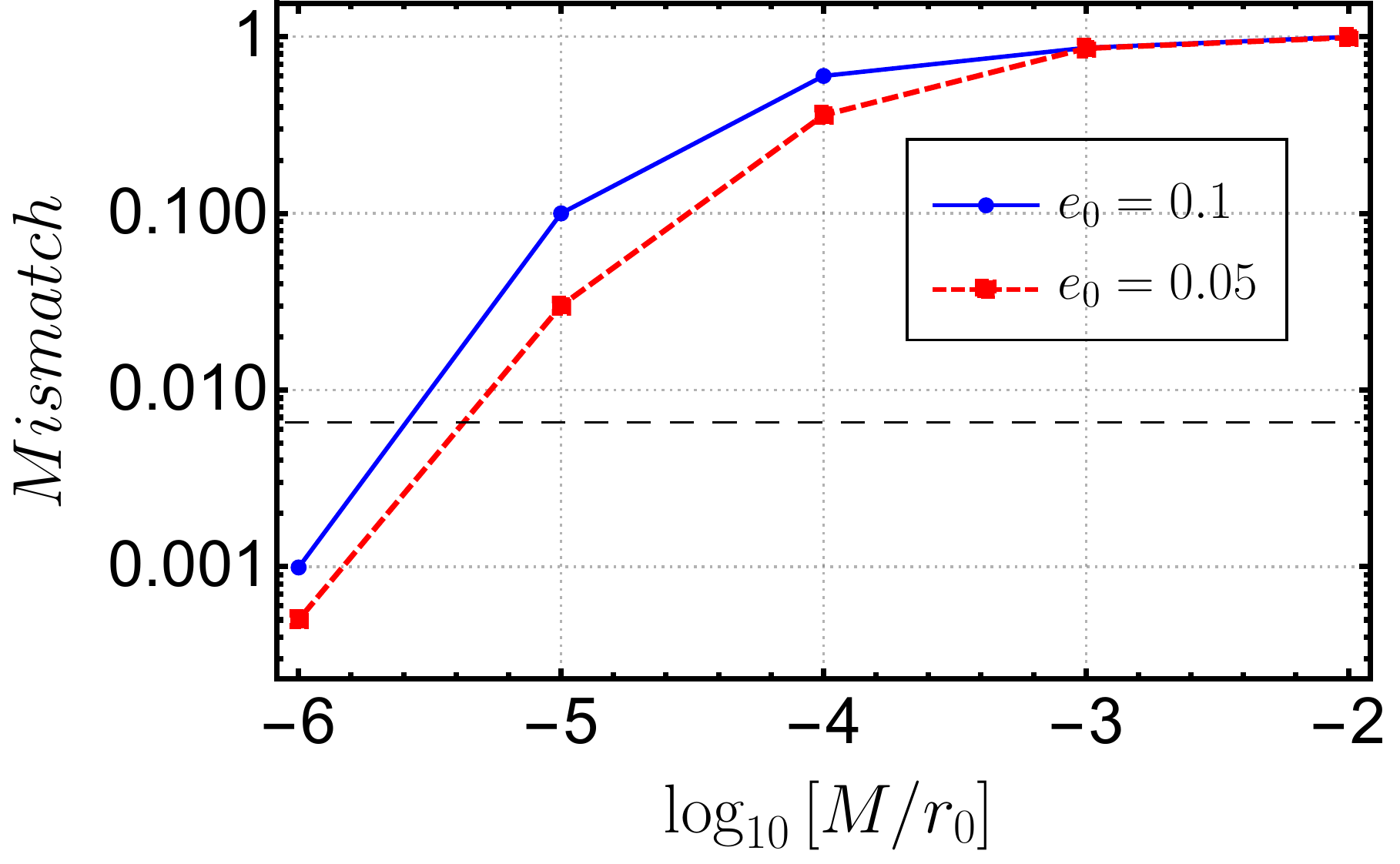}
   \caption{
   The results of the mismatch between GW waveforms with and without DM halos for different compactness $M/r_0$ and initial eccentricity $e_0$.
   The black dashed line corresponds to the threshold $d/(2\,\text{SNR}^2)\approx 0.007$.
   }
    \label{fig:3.05}
\end{figure}

To distinguish the waveforms more accurately, we calculate the mismatch between GW signals emitted from EMRIs with and without DM halos.
Given two signals $h_1(t)$ and $h_2(t)$, the inner product $(h_1|h_2)$ is defined as
\begin{equation}\label{overlap}
    (h_1|h_2)=2\int_0^{+\infty } \frac{\tilde{h}_1(f)\tilde{h}_2^*(f)+ \tilde{h}_2(f)\tilde{h}_1^*(f)}{ S_n(f)}\,df,
\end{equation}
where $\tilde{h}(f)$ is the Fourier transformation of the time-domain signal $h(t)$, $\tilde{h}^*$ denotes the complex conjugate of $\tilde{h}$, and the SNR for the signal $h$ is $\rho=\sqrt{(h|h)}$.
For LISA, the one-side noise power spectral density is \cite{Robson:2018ifk}
\begin{equation}
\begin{split}
    \label{psd-lisa}
    S_n(f) =&\frac{S_x}{L^2}+\frac{2S_a \left[1+\cos^2(2\pi\,f L/c)\right]}{(2\,\pi f)^4 L^2}\\
    &\times\left[1+\left(\frac{4\times 10^{-4}\text{Hz}}{f}\right) \right],
\end{split}
\end{equation}
where $\sqrt{S_a}=3\times 10^{-15}\ \text{m s}^{-2}/\text{Hz}^{1/2}$ is the acceleration noise, $\sqrt{S_x}=1.5\times 10^{-11}\ \text{m/Hz}^{1/2}$ is the displacement noise, and $L=2.5 \times 10^6\text{ km}$ is the arm length of LISA \cite{LISA:2017pwj}.
The overlap between two GW signals is \cite{Babak:2006uv}
\begin{equation}
    \mathcal{O}(\tilde{h}_1,\tilde{h}_2)= \frac{(\tilde{h}_1|\tilde{h}_2)}{\sqrt{(\tilde{h}_1|\tilde{h}_1)(\tilde{h}_2|\tilde{h}_2)}},
\end{equation}
and the mismatch between two signals is defined as
\begin{equation}\label{mismatch}
    \text{Mismatch}=1-\mathcal{O}_\text{max}(\tilde{h}_1,\tilde{h}_2),
\end{equation}
where the maximum is evaluated with respect to time and phase shifts.
The mismatch is zero if two signals are identical.
Two signals are considered experimentally distinguishable if their mismatch is larger than $d/(2\,\text{SNR}^2)$ \cite{Flanagan:1997kp,Lindblom:2008cm}, where $d=13$ is the number of intrinsic parameters of the GW source.
Considering EMRIs with masses $(10^6+10)M_\odot$ at $d_L=1$ Gpc and the integration time of one year before the coalescence, 
we calculate the mismatch between GW waveforms with and without DM halos and the results with LISA are shown in Fig. \ref{fig:3.05}. 
The SNR is about 32 for the GW signals from EMRIs considered above.
The initial eccentricity $e_0$ is chosen at $p_0=6.3M_\text{BH}$.
As shown in Fig. \ref{fig:3.05}, if the compactness of the DM halo $M/r_0$ is larger, then the mismatch between GW waveforms with and without DM halos is bigger, 
i.e., more compact DM halos can be detected more easily with LISA.
Again, eccentric orbits can detect smaller compactness.
Therefore, we can use GWs from EMRIs within the environments of galaxies to test the existence of DM halos and detect the compactness of halos with $M/r_0$ as small as $10^{-5}$.

\section{Parameter Estimation for Circular EMRIs}
\label{sec:fim}

To study degeneracies among the source parameters and assess the detector's capacity to constrain the compactness, we perform parameter estimations with the FIM method \cite{Cutler:1994ys,Poisson:1995ef}.

For the GW signal $h$, the FIM is defined as
\begin{equation}\label{FIM}
    \Gamma_{ij}=\left.\left(\frac{\partial h}{\partial \xi^i} \right | \frac{\partial h}{\partial \xi^j} \right),
    \end{equation}
where $\xi^i$ denotes the source parameter.
The statistical error on $\xi^i$ is
\begin{equation}
\label{error-FIMa}
\sigma_{i}=\Sigma_{i i}^{1 / 2},
\end{equation}
and the correlation coefficient between the parameters $\xi^i$ and $\xi^j$ is
\begin{equation}
\label{error-FIMb}
c_{ij}=\Sigma_{i j} /\left(\sigma_{i} \sigma_{j}\right),
\end{equation}
where $\Sigma_{ij}=(\Gamma^{-1})_{ij}$ is the inverse of the FIM.
Because of the triangle configuration of the space-based GW detector,
the detector can be regarded as a network of two L-shaped detectors \cite{Cutler:1997ta}, 
so the total SNR is the sum of the SNRs of two L-shaped detectors,
\begin{equation}
\rho=\sqrt{\rho_1^2+\rho_2^2}=\sqrt{\left( s_1|s_1 \right)+\left( s_2|s_2 \right)},
\end{equation}
where $s_1$ and $s_2$ denote the signals detected by the two L-shaped detectors, respectively.
The total covariance matrix of the source parameters is obtained by inverting the sum of the Fisher matrices,
and the parameter error is
\begin{equation}
\label{2error-FMI}
 \sigma_i^2=(\Gamma_1+\Gamma_2)^{-1}_{ii}.   
\end{equation}

Taking the Newtonian gravity and dynamical friction of DM halos into account, 
analytical waveforms of EMRIs with DM halos in quasi-circular orbits were derived in \cite{Eda:2014kra}
under the stationary phase approximation \cite{Cutler:1994ys,Will:1994fb}.
Following the approach in \cite{Maselli:2021men,Zhang:2024ugv}, 
we use the FIM method to estimate the parameter error numerically.
The GW waveforms of EMRIs in quasi-circular orbits are
\begin{equation}
\begin{split}
h_+=&\frac{4\mu(t)\Omega(t)^{2/3}M_\text{BH}^{2/3}}{d_L}\frac{1+\cos^2\iota}{2}\cos\left[2\varphi(t)\right]\\
&\times\left(1-\frac{4M}{3r_0}\right),
\end{split}
\end{equation}
\begin{equation}
\begin{split}
h_\times=&\frac{4\mu(t)\Omega(t)^{2/3}M_\text{BH}^{2/3}}{d_L}\cos\iota\sin\left[2\varphi(t)\right]\\
&\times\left(1-\frac{4M}{3r_0}\right),
\end{split}
\end{equation} 
where $\Omega(t)=d\varphi/dt$ is the orbital frequency, 
the orbital phase $\varphi(t)$ is 
\begin{equation}
    \label{verphi-t}
    \varphi(t)=\varphi_c+\int_0^t\Omega(t)dt, 
\end{equation}
and $\varphi_c$ is the initial orbital phase at $t=0$.
Combining Eqs. \eqref{dphidc}, \eqref{dtdc}, \eqref{dE-B}, \eqref{dL-B} and \eqref{verphi-t} for $e_0=0$, we can obtain $\mu(t)$, $\Omega(t)$, and $\varphi(t)$ numerically.  

The GW strain measured by the detector is
\begin{equation}\label{ht-signal}
h(t)=h_{+}(t) F^{+}(t)+h_{\times}(t) F^{\times}(t),
\end{equation}
where the interferometer pattern functions $F^{+}(t)$ and $F^{\times}(t)$ can be expressed in terms of the source orientation $(\theta_s,\phi_s)$ and the direction of the angular momentum $(\theta_1,\phi_1)$.
Because of the orientation dependence, GW signals are modulated due to the orbital motion \cite{Babak:2006uv, Maselli:2021men}.
Removing the extrinsic parameters $\{\,\theta_s, \phi_s, \theta_1, \phi_1, d_L\,\}$,
the source parameters related to the evolution of EMRIs within DM halos are
\begin{equation}
    \hat{\bm{\xi}}=\{\, \ln M_{\rm{BH}}, \ln \mu_0,\ln M_{\rm{halo}}, \ln r_0, r_c, \varphi_c\,\},
\end{equation} 
where $\mu_0$, $r_c$, and $\varphi_c$ are the mass of the SCO, the orbital radius and the orbital phase at $t=0$, respectively.
Combining Eqs. \eqref{FIM}, \eqref{error-FIMa}, \eqref{2error-FMI}, and \eqref{ht-signal},
we estimate the errors of the parameters $\hat{\bm{\xi}}$.

\begin{figure*}[htp]
    \centering
    \includegraphics[width=0.99\linewidth]{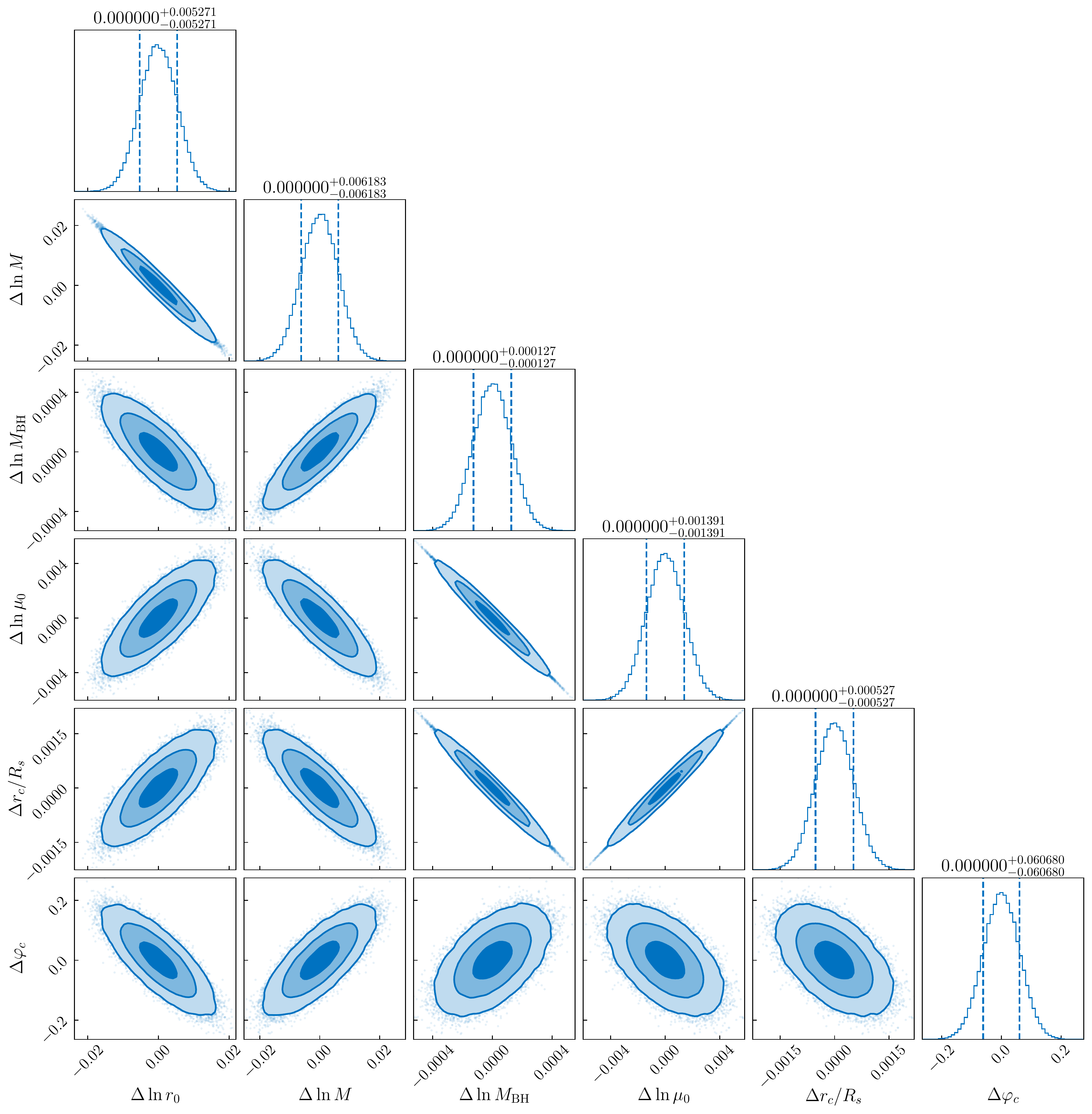}
    \caption{Corner plot for the probability distribution of the source parameters, inferred after one-year observations with LISA.
    The halo parameters $M/r_0$ and $M/r^2_0$ are $10^{-1}$ and $10^{-6}/M_\text{BH}$.
    Vertical lines show the $1\sigma$ interval for each waveform parameter. 
    The contours correspond to the $68\%$, $95\%$, and $99\%$ probability confidence intervals.}
    \label{DMcorner01}
\end{figure*}

\begin{figure*}[htp]
    \centering
    \includegraphics[width=0.99\linewidth]{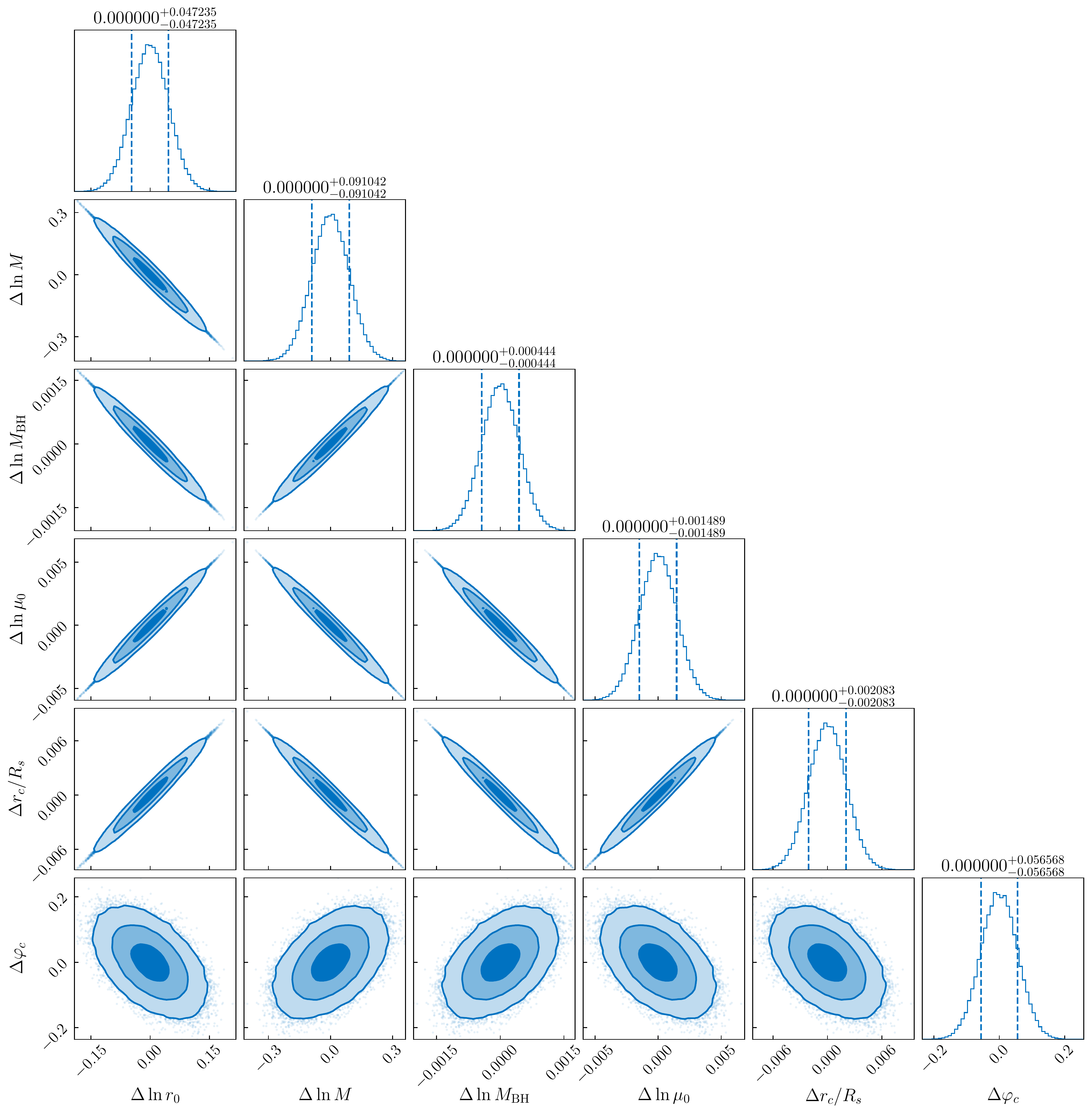}
    \caption{Corner plot for the probability distribution of the source parameters, inferred after one-year observations with LISA.
    The halo parameters $M/r_0$ and $M/r^2_0$ are $10^{-2}$ and $10^{-6}/M_\text{BH}$.
    Vertical lines show the $1\sigma$ interval for each waveform parameter. 
    The contours correspond to the $68\%$, $95\%$, and $99\%$ probability confidence intervals.}
    \label{DMcorner02}
\end{figure*}

\begin{figure*}[htp]
    \centering
    \includegraphics[width=0.99\linewidth]{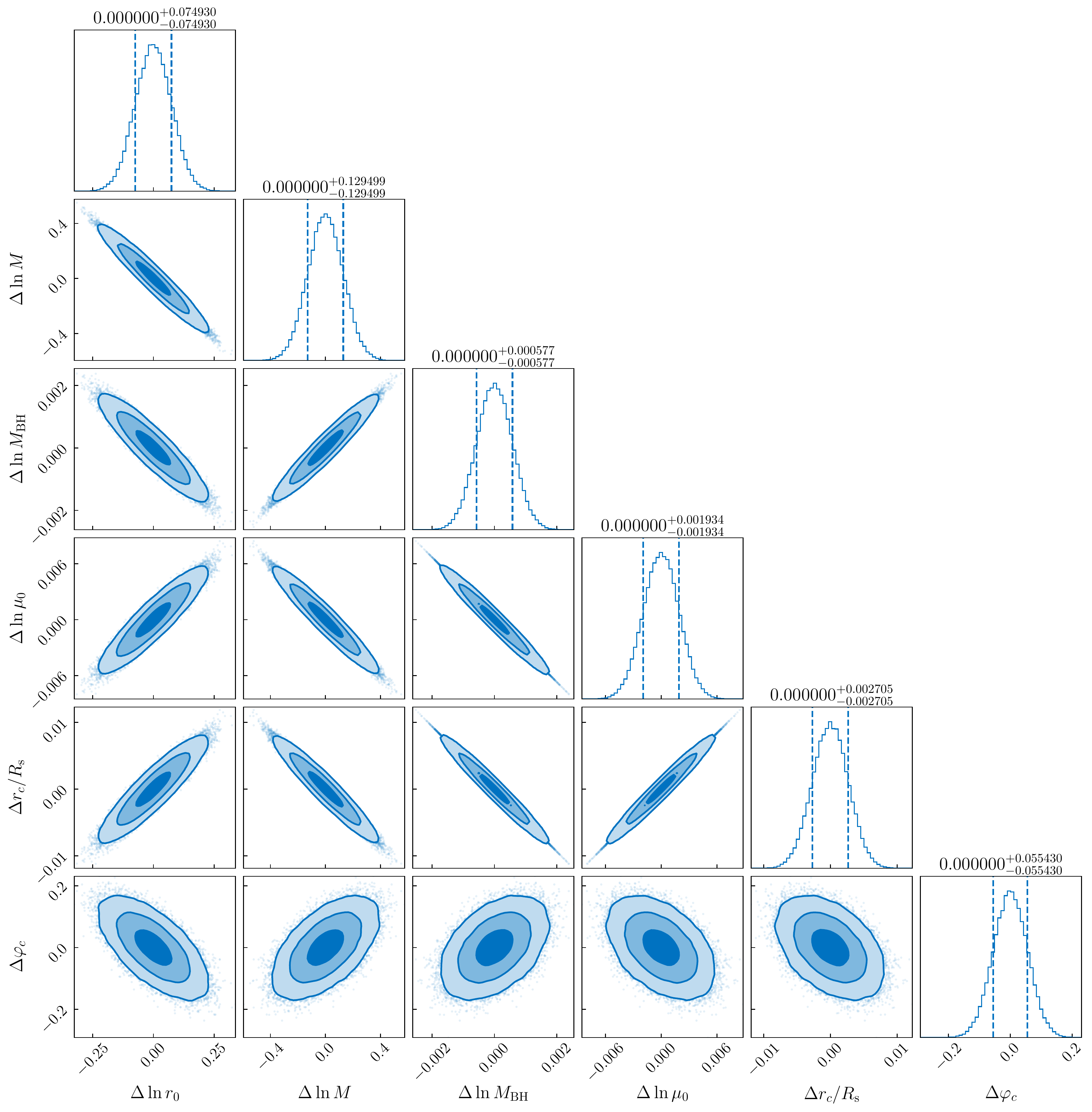}
    \caption{Corner plot for the probability distribution of the source parameters, inferred after one-year observations with LISA.
     The halo parameters $M/r_0$ and $M/r^2_0$ are $10^{-2}$ and $10^{-7}/M_\text{BH}$.
    Vertical lines show the $1\sigma$ interval for each waveform parameter. 
    The contours correspond to the $68\%$, $95\%$, and $99\%$ probability confidence intervals.}
    \label{DMcorner03}
\end{figure*}

We take one year of observation time before the ISCO and fix the SNR to be $\rho=30$ \cite{Berti:2015itd,Maselli:2021men}. 
The source and orbital orientations are fixed as $\theta_s=\pi/3$, $\phi_s=\pi/2$, $\theta_1=\pi/4$ and $\phi_1=\pi/4$. The initial orbital phase $\varphi_c=0$, and $r_c$ is adjusted to experience one-year adiabatic evolution before the ISCO.
For LISA, the lower and upper cutoff frequencies are $f_\text{low}=10^{-4}$ Hz and $f_\text{high}=1$ Hz \cite{Yagi:2009zm},
so the lower and upper limits of the integration
are chosen as $f_\text{ini}=\text{Max}(f_\text{low}, f_\text{1yr})$ and $f_\text{end}=\text{Min}(f_\text{high}, f_\text{ISCO})$, respectively. 
$f_\text{1yr}$ is the GW frequency at one year before the ISCO, 
and $f_\text{ISCO}$ is the frequency at the ISCO. 
We show the estimated parameter errors of EMRIs with the numerical waveforms in Table \ref{tab:3.3}.
The probability distribution obtained from the FIM approach for the parameters of EMRIs are shown in Figs. \ref{DMcorner01}, \ref{DMcorner02}, and \ref{DMcorner03}.

\begin{widetext}

\begin{table}[htbp!]
\centering
\begin{tabular}{ p{1.7cm} p{1.6cm} p{2.0cm} p{1.8cm} p{1.8cm} p{1.5cm} |p{1.4cm} p{2.0cm} }
    \hline
    $\sigma(\ln r_0)$   & $\sigma(\ln M)$   & $\sigma (\ln M_\text{BH})$  & $\sigma(\ln \mu_0)$ & $\sigma(r_c)/R_s$ & $\sigma(\varphi_c)$  & $M/r_0$  & $M/r^2_0$ ($M_\text{BH}^{-1}$)\\
    \hline
    $0.00527$  & $0.00618$   & $0.000127$   & $0.00139$   & $0.000527$   & $0.0607$    & $10^{-1}$    & $10^{-6}$ \\
    $0.0472$   & $0.0910$    & $0.000444$   & $0.00149$   & $0.00208$   & $0.0566$    & $10^{-2}$    & $10^{-6}$ \\
    $0.0749$    & $0.129$    & $0.000577$   & $0.00193$   & $0.00270$   & $0.0554$    & $10^{-2}$    & $10^{-7}$ \\    
    \hline
    \hline
\end{tabular}
    \caption{
The estimated errors of the parameters of GW signals from EMRIs within DM halos with LISA.
The mass of the central MBH is $10^6M_\odot$, and the mass of the small BH is $10M_\odot$.
The DM halo parameters $M/r_0$ and $M/r_0^2$ are listed in the last two columns. 
The halo parameters $(M, r_0)$ for the three rows are $(10^4M_\text{BH}, 10^5M_\text{BH})$, $(10^2M_\text{BH}, 10^4M_\text{BH})$ and $(10^3M_\text{BH}, 10^5M_\text{BH})$, respectively.
    }
\label{tab:3.3}
\end{table}

\end{widetext}

Comparing the results in the second and third rows in Table \ref{tab:3.3},
we see that the estimated errors of the parameters (especially $\sigma(\ln r_0)$ and $\sigma(\ln M)$) decrease as the halo density $M/r^2_0$ increases when the halo compactness remains the same.
Comparing the results in the first and second rows in Table \ref{tab:3.3},
we see that the increase of the halo compactness $M/r_0$ can make the error of the parameters decrease when $M/r^2_0$ remains the same.
When $M/r_0$ changes from $10^{-2}$ to $10^{-1}$, the errors of 
$\ln M$ and $\ln r_0$ are reduced by almost an order of magnitude.
Therefore, larger halo compactness $M/r_0$ and halo density $M/r^2_0$ decrease the estimated errors of the parameters and improve the detection accuracy for DM halos.
Comparing Figs. \ref{DMcorner01} and \ref{DMcorner02}, 
we find that the correlations between $\ln{r_0}$ and the other parameters, as well as those between $\ln{M}$ and the other parameters, decrease when $M/r_0$ changes from $10^{-2}$ to $10^{-1}$.
The differences in the correlations between the parameters in Figs. \ref{DMcorner02} and  \ref{DMcorner03} are not significant, 
probably because the dependence on the halo density $M/r_0^2$ is not strong.
Therefore, larger halo compactness is likely to reduce the correlations between the parameters and help to break the degeneracy between the parameters.

\section{Conclusions and Discussions}
\label{conclusion}
Using the analytic, static, and spherically symmetric metric for a Schwarzschild BH immersed in DM halos with the Hernquist-type density distribution, 
we derive analytic formulas for the orbital period and orbital precession of eccentric EMRIs within the environment of DM halos. 
The results show that the presence of the DM halo retards the orbital precession and even retrogrades the orbital precession if the local density of DM halos $\rho_\text{DM}\sim M/r_0^2$ is sufficiently large.
For larger orbits,  
the retrograde precessions decrease faster in the presence of DM halos.
With DM halos, the retrograde-to-prograde precession transition happens at some critical value of $p$ and then the prograde precessions increase as $p$ continues to decrease. 

Taking the dynamical friction, accretion, and GW reaction into account, 
we calculate the corresponding energy and angular momentum fluxes. 
We find that the effect of the GW reaction is significantly greater than those of dynamical friction and accretion when $p<15M_\text{BH}$. 
Furthermore, we carry out numerical calculations to study the evolution of EMRIs under the combined influences of the dynamical friction, accretion and GW reaction.
We find that EMRIs immersed in DM halos evolve more slowly than those without DM.
Comparing the numbers of orbital cycles with and without DM halos accumulated over one-year evolution before the merger, 
we find that DM halos with the compactness as small as $10^{-5}$ can be detected.
By calculating the mismatch between GW waveforms with and without DM halos,
we show that we can use GWs from EMRIs within the environments of galaxies to test the existence of DM halos and detect the compactness as small as $10^{-5}$.
We also find that eccentric orbits can help to detect DM halos with smaller compactness.
Considering the degeneracies among the source parameters, 
we estimated the parameter errors with the FIM method numerically.
We find that larger values of halo compactness and density reduce the estimated errors of the parameters,
and larger halo compactness helps to break the degeneracy between the parameters.

To further improve the accuracy of the GW waveform of EMRIs within DM halos, 
there is still a lot of work to be done. 
The forms of DF and accretion we used are based on Newtonian estimated, and the relativistic effects of DF and accretion are not considered in this paper. 
Many studies have explored the relativistic effects of DF and accretion \cite{Speeney:2022ryg,Vicente:2022ivh,Traykova:2023qyv,Mach:2021zqe,Hughes:2018qxz}. 
Incorporating the relativistic effects of DF and accretion from first principles into the cases of EMRIs with DM halos deserves further attention. 
Compared to the perturbation black hole approach, the quadrupole formula for the energy and angular momentum fluxes of GW used in this paper is less accurate, 
especially for highly eccentric orbits where higher-order multipoles are important. 
Currently, we are focusing on using the perturbation BH approach to study the evolution of EMRIs in DM halos, taking DF and accretion into account.
We also need to consider the DM halo feedback \cite{Kavanagh:2020cfn,Becker:2021ivq,Karydas:2024fcn}.
To distinguish the effect of DM halos on GWs from other mediums (e.g., accretion disks) or modified gravity \cite{Cardoso:2019rou, Becker:2022wlo, Zhang:2022hbt, Cardoso:2018zhm, Maselli:2020zgv},
further investigation is deserved.

\begin{acknowledgments}
The computing work in this paper is supported by the Public Service Platform of High Performance Computing by Network and Computing Center of HUST.
This research is supported in part by the National Key Research and Development Program of China under Grant No. 2020YFC2201504.
\end{acknowledgments}

\end{CJK*}


%

\end{document}